\documentclass[doublecol]{epl2}

\title{Experimental constraints on anti-gravity and antimatter, in the context of dark energy}
\shorttitle{Experimental constraints on anti-gravity and antimatter}

\author{Yuan-Sen Ting\inst{1}}
\shortauthor{Y.-S. Ting}

\institute{\inst{1} Harvard--Smithsonian Center for Astrophysics, 60 Garden Street, Cambridge, MA 02139, USA}


\abstract{In a paper by Villata (2011) \cite{vil11}, the possibility of a repulsive gravitational interaction between antimatter and ordinary matter was discussed. The author argued that this anti-gravity can be regarded as a prediction of general relativity, under the assumption of CPT symmetry. Stringent experimental constraints have been established against such a suggestion. The measurement of free-fall accelerations of various nuclei by the E\"ot-Wash group and searches for equivalence principle violation through the gravitational splitting in kaon physics consistently establish null results on any difference between the gravitational behaviour of antimatter and ordinary matter. The original arguments against antigravity were questioned by Nieto \& Goldman (1991) \cite{nie91}. In the light of new experiments as well as theoretical developments in the past 20 years, some of Nieto \& Goldman's concerns have been addressed. While a precise measurement of the free-fall acceleration of antihydrogen will eventually lay this issue to rest, the purpose of this short letter is to argue that the substitution of dark energy with anti-gravity by antimatter, as suggested by Villata, is highly unlikely.}

\begin{document}

\maketitle

%
%
%
%
%
%
\section{Introduction}\label{sec:introduction}

Since the initial discovery of the accelerating expansion of the Universe \cite{rie98,per99}, subsequent measurements, including baryon acoustic oscillation from SDSS\footnote{Sloan Digital Sky Survey} \cite{and12}, WiggleZ \cite{bla11}, 6dFGS\footnote{Six-degree Field Galaxy Survey} \cite{beu11}, type Ia supernovae \cite{rie11} and Cosmic Microwave Background anisotropy \cite{ade13} all support the dark energy scenario. For example the most recent study from the Plank satellite \cite{ade13}, shows $\Omega_\Lambda \sim 0.7$ in a $\Lambda$CDM model. However, the nature of dark energy remains poorly understood. Theoretical attempts are far from successful, leaving dark energy one of the biggest mysteries in present day scientific research.

Villata (2011) \cite{vil11} argues that when CPT symmetry is applied to general relativity, this predicts a repulsive gravitational interaction between matter and antimatter. Moreover, this paper presents a scenario where the accelerating expansion of the Universe is driven by this anti-gravity behaviour and the antimatter could be residing in the large scale (Mpc scale) voids of ordinary matter.

This might seem to be an attractive alternative since it is difficult to make direct tests of the gravitational properties of antimatter. The experimental difficulties include (a) the weakness of the gravitational effect compared to the other interactions such as the electromagnetic interaction; the difficulty of reducing the electromagnetic influences to a negligible level, and (b) the propensity of antimatter to annihilate when it comes into contact with ordinary matter. 

However, such a claim is not unprecedented (for a historical review, see \cite{nie91}). Although there is still no conclusive direct experimental measurement (but see \cite{cha13} for preliminary results), and pending relevant data from AEGIS\footnote{Antihydrogen Experiment: Gravity, Interferometry, Spectroscopy} \cite{dob07}, ATRAP\footnote{The Antihydrogen Trap} \cite{gab12} and ALPHA\footnote{Antihydrogen Laser Physics Apparatus} \cite{cha13} experiments at CERN, indirect measurements of equivalence principle violation have already put stringent and quantifiable constraints on the anti-gravity behaviour of antimatter, essentially denying such a possibility.

Objections were raised against Villata's (2011) \cite{vil11} suggestion from a theoretical point of view\cite{cro11,cab12}. While Villata (2012) \cite{vil12} attempted to address these theoretical objections, this letter focuses on experimental and observational constraints.  

Original arguments against antigravity were first proposed by Morrison (1958) \cite{mor58}, Schiff (1959) \cite{sci59} and Good (1961) \cite{goo61}. Some of these arguments were discussed and challenged by Nieto \& Goldman (1991) \cite{nie91}. It is important to note that, while it is impossible to exhaust all theoretical models as pointed out by Nieto \& Goldman, persistent experimentalists have taken Nieto \& Goldman seriously and have addressed some of their concerns.

In the next section, we will discuss the E\"ot-Wash group's experiment. We then discuss the study of matter and antimatter gravitational coupling based on kaon physics. Both experiments show null results and the precisions they achieve already exceed the predicted signal from anomalous antimatter gravitational behaviour. We will conclude in the final section by extending these baryonic results also to the dark matter sector.

%
%
%
%
%
%
\section{E\"ot-Wash group's experiment}\label{sec:torsion}

The E\"ot-Wash group's experiment from the University of Washington uses a torsion balance (for experimental details, refer to \cite{wag12}). They measure the E\"otv\"os parameter \cite{eot22} which is the difference in free-fall acceleration ${\bf a_1}$ and ${\bf a_2}$ between two materials, 
\begin{equation}
\eta = \frac{|{\bf a_1}|-|{\bf a_2}|}{(|{\bf a_1}|+|{\bf a_2}|)/2},
\end{equation}

\noindent
under an external gravitational field. Note that under a constant gravitational field, the free-fall acceleration is proportional to the ratio of gravitational mass $m_g$ to inertial mass $m_i$. Therefore, the $\eta$ parameter can be regarded as the difference in this ratio,
\begin{equation}
\eta = \frac{m_{g,1}/m_{i,1} - m_{g,2}/m_{i,2}}{(m_{g,1}/m_{i,1}+m_{g,2}/m_{i,2})/2}.
\end{equation}

Before we move on, we pause to define gravitational mass and inertial mass carefully which will be helpful in clarifying our further discussions. Gravitational mass refers to the proportionality constant of the force exerted by an external gravitational field on an object. Whereas the inertial mass designates the ratio of any force to the acceleration of an object. Let $m_i$, $m_{\bar{i}}$, $m_g$ and $m_{\bar{g}}$ be the inertial mass and gravitational mass of ordinary matter and antimatter, respectively. Assuming CPT symmetry (also confirmed to at least $10^{-10}$ precision in the charge to inertial mass ratio in a proton-antiproton study \cite{gab99}), implies $m_i = m_{\bar{i}}$. The E\"ot-Wash results are sensitive to any difference between $m_i/m_g$ and $m_{\bar{i}}/m_{\bar{g}}$. Since any deviation between $m_i$ and $m_g$ can be absorbed into the gravitational constant, E\"ot-Wash is essentially measuring the deviation of $m_{\bar{g}}$ from $m_{\bar{i}}$, and hence is probing the difference in $m_{\bar{g}}$ and $m_g$.

The purpose of measuring the E\"otv\"os parameter is to detect any anomalies that could deviate the effective gravitational mass from the inertial mass. Note that these effects could vary across nuclear species, {\it e.g.} the amounts of virtual particle contents are different for different nuclear species. If there are anomalies or new gravitational interactions between matter-matter or matter-antimatter, this difference across the periodic table manifests itself in the E\"otv\"os parameter for different nuclei species. In other words, if the detected upper limit of E\"otv\"os parameter is much smaller than the predicted value from the anomalies, one can constrain the predicted anomalies with a quantifiable bound.

Nieto \& Goldman (1991) \cite{nie91} first raised concerns on using the E\"otv\"os parameter to constrain deviation of antimatter gravitational behaviour. They mentioned that ruling out a certain model parameter space ({\it e.g.} study from Adelberger {\it et al.} 1991 \cite{ade91}) is not sufficient to conclude that antimatter gravitates normally-- there is just too much freedom to construct different types of models. While this objection might be valid, one can argue more generally using arguments first pointed out by Schiff (1959) \cite{sci59}. The underlying idea is that although positrons, unlike electrons, cannot exist as stable particles in atoms, they do exist as virtual particles. For example, the polarization of the vacuum due to the Coulomb field between electrons and nucleus is supported by the agreement between the measured and calculated value of the Lamb Shift \cite{gum05}. More importantly, the magnitude of this effect depends on the atomic number. In other words, the probability of having virtual positrons (and hence antimatter) varies from atom to atom, and so does their contribution to the anomalies in gravitational mass.
 
Now consider the case where antimatter gravitates differently from matter. The challenge is to estimate the effect anti-gravity might have on the $m_g/m_i$ ratio, for different elements. Nieto \& Goodman pointed out that the estimate from Schiff (1959) \cite{sci59} is questionable as Schiff based his calculation on a non-renormalizable system. They argued that the signal in the E\"otv\"os parameter might be as tiny as $10^{-16}$ and therefore evades detection. A more recent paper by Alves, Jankowiak \& Saraswat (2009) \cite{alv09}, calculates that if antimatter were to gravitate differently from matter, the effect of antiquarks in the nucleon should have already been detected with E\"ot-Wash's experiment. The Alves {\it et al.} (2009) calculation constrains the fractional deviation of $g_H$ and $g_{\bar{H}}$ to be less than $10^{-9}$, where $g_H$ and $g_{\bar{H}}$ are the free-fall accelerations of hydrogen and antihydrogen. The bottom line is that the difference between $g_H$ and $g_{\bar{H}}$ cannot be $2g_H$ as claimed by Villata's theory.

%
%
%
%
%
%
\section{Constraint on equivalence principle through kaon physics}\label{sec:kaon}

The idea of probing anomalous antimatter gravitational behaviour using neutral kaons was first proposed in Good (1961) \cite{goo61} and was discussed in Nieto \& Goodman (1991) \cite{nie91}. The phases of the perturbed wavepackets of matter and antimatter diverge if they experience different gravity. This in turn will perturb the delicate oscillation observations in the $K_0 \bar{K}_0$ system. In fact, stringent upper limits on such a violation have been established \cite{ham98}. Hambye, Mann \& Sarkar (1998) \cite{ham98} attempted to quantify this violation. Since $K^0$ and $\bar{K}^0$ form a matter-antimatter pair, such a study is directly addressing the possibility of ordinary matter and antimatter having different gravitational behaviour.

Although as pointed out by Nieto \& Goodman (1991) \cite{nie91}, this original idea of Good assumes an absolute gravitational potential that could not have physical meaning in this context, one can nevertheless assume a gravitational field due to the earth and study the question of how fine-tuned the absolute zero of the potential has to be in order to evade the experiment. Based on the kaon physics measurement and assuming CPT, Hambye {\it et al.} (1998) showed that if we assume that the kaons $K_0$ and $\bar{K}_0$ have an energy difference of $2mV$ in the local potential, as claimed by Villata, where $m$ is the mass of kaons and $V$ the local gravitational potential, the kaon oscillation observations require a fine-tuning to set the absolute zero of the gravitational potential to within $10^{-12}$ of $Gm_e/R_e$, where $m_e$ and $R_e$ are the earth mass and earth radius, respectively. This unlikely fine-tuning strongly disfavours anti-gravitational behaviour of antimatter.

%
%
%
%
%
%
\section{Extension to the dark matter sector and conclusion}\label{sec:conclusion}

Based on the results from torsion-balance and kaon experiments, one can put severe constraints on anti-gravity behaviour between anti-matter and matter, at least in the baryonic sector, essentially denying Villata's (2011) claim. Direct gravitational experiments using antihydrogen can give a definitive answer to this issue. In fact preliminary results have been published recently \cite{cha13}, however the precision is rather weak (fractional deviation is only constrained to the order of $10^2$) at this point.

So far, we only showed that baryonic matter and anti-matter cannot anti-gravitate toward each other. However, we know that the large scale structure in the cosmos and the accelerating expansion of the Universe in the $\Lambda$CDM model is dominated by dark matter and dark energy. In the context of Villata's suggestion, dark energy is excluded, the gravitational behaviour of dark matter determines the evolution of the Universe. 

Gradwohl \& Frieman (1992) \cite{gra92} showed that if there is an additional long range force within the dark matter sector, as in the case of Villata's suggestion and dark matter anti-gravitates with anti dark matter, large-scale structure and clustering will change drastically. This violates the agreement of large-scale structure predictions and observations in cosmology such as the baryon acoustic oscillation. Therefore anti-gravitational behaviour in between equal amount of dark matter and dark antimatter was ruled out.

\acknowledgments
YST thanks Christopher W. Stubbs for helpful discussions and comments. YST is supported by the Harvard University Faculty of Arts and Science through a graduate school scholarship and research assistantship.

\end{document}